\documentclass[preprint2]{aastex}

\shorttitle{Kinematics \& Chemistry of Halo Substructures: The Vicinity of the VOD}
\shortauthors{Casey, Keller \& Da Costa}

\begin{document}

\title{Kinematics \& Chemistry of Halo Substructures: The Vicinity of the Virgo Over-Density}

\author{Andrew R. Casey, Stefan C. Keller and Gary Da Costa}
\affil{Research School of Astronomy \& Astrophysics,
	Australian National University, Mount Stromlo Observatory, via Cotter Rd, Weston, ACT 2611, Australia}

\begin{abstract}
We present observations obtained with the AAT's 2dF wide field spectrograph AAOmega of K-type stars located within a region of the sky which contains the Virgo Over-Density and the leading arm of the Sagittarius Stream.  On the basis of the resulting velocity histogram we isolate halo substructures in these overlapping regions including Sagittarius and previously discovered Virgo groups. Through comparisons with \textit{N}-body models of the Galaxy-Sagittarius interaction, we find a tri-axial dark matter halo is favoured and we exclude a prolate shape. This result is contradictory with other observations along the Sagittarius leading arm, which typically favour prolate models.  We have also uncovered K-giant members of Sagittarius that are notably more metal poor ($\langle[$Fe/H$]\rangle = -1.7\,\pm\,0.3$ dex) than previous studies. This suggests a significantly wider metallicity distribution exists in the Sagittarius Stream than formerly considered. We also present data on five carbon stars which were discovered in our sample.
\end{abstract}

\keywords{Galaxy: halo, structure --- Individual: Sagittarius, Virgo Stellar Stream --- Stars: carbon, K-giants}

\section{Introduction}
\label{sec:introduction}

The proportion of substructure recently uncovered in the Galaxy has highlighted the crucial involvement accretion has played in the formation of the Milky Way \citep[see][and references therein]{Helmi_2008}. Properties of these stellar structures allow us to probe the formation mechanisms and history of the Galaxy. Recent studies \citep{Carollo;et-al_2007, Carollo;et-al_2010} have suggested multiple evolutionary paths are required for galaxy formation to reconcile observational evidence, although this is a subject of ongoing debate \citep{Schoenrich;et-al_2011}. Regardless, the dissipation-less merging paradigm is widely accepted and consistent with favoured Cold Dark Matter ($\Lambda$CDM) cosmology models. Through the examination of ongoing accretion events in the Milky Way and fossils from previous mergers we can trace the evolution of the outer most regions of the Galaxy \citep[e.g.,][]{Helmi;White_2001}.
	
 Accretion is at least partly \citep[e.g.,][]{Starkenburg;et-al_2009}, if not entirely responsible for the formation of the stellar halo. \citet{Bell;et-al_2008} compared Sloan Digital Sky Survey \citep[hereafter SDSS]{York;et-al_2000} data to galaxy formation simulations using different dark halos and found that observations are consistent with the stellar halo being entirely formed by hierarchical merging of accreted satellites (see also \citet{Xue;et-al_2011}).
	
Unquestionably the most prominent ongoing accretion event within the Milky Way is that of the Sagittarius (Sgr) dwarf Spheroidal (dSph) galaxy. Originally discovered by \citet{Ibata;et-al_1994} as a co-moving group of K- and M-type giants, the tidal tails of Sgr circle our Galaxy. As such they have been extensively traced with red-clump stars \citep{Majewski;et-al_1999}, carbon stars \citep{Totten;Irwin_1998, Ibata;et-al_2001}, RR Lyrae stars \citep{Ivezic;et-al_2000, Vivas;et-al_2005, Keller;et-al_2008, Watkins;et-al_2009, Prior;et-al_2009b}, A-type stars \citep{Newberg;et-al_2003}, BHB stars \citep{Ruhland;et-al_2011} and K/M-giants \citep{Majewski;et-al_2003,  Yanny;et-al_2009, Keller;Yong;Da_Costa_2010}. Tracers originating from the host system can be unequivocally identified with spatial and kinematical information because they remain dynamically cold, and are identifiable as kinematic substructures long after they are stripped from their progenitor \citep[for example][]{Ibata;Lewis_1998, Helmi;White_1999}. 
	
Stellar tracers within these tidal tails are kinematically sensitive to the galactic potential. This has led various groups to model the Sgr interaction with different dark matter profiles. \citet{Martinez-Delgado;et-al_2004} traced the Northern leading arm and found a near spherical or oblate ($q \approx 0.85$) dark matter halo best represented the observed debris, coinciding with the findings of \citet{Ibata;et-al_2001}. In contrast, \citet{Helmi_2004} found evidence in the Sgr leading debris that most favoured a prolate ($q = 1.25$) halo. \citet{Vivas;et-al_2005} found that either a prolate or spherical model of  \citet{Helmi_2004} would fit their RR Lyrae observations, rather than those of an oblate model. \citet{Johnston;et-al_2005} later pointed out that no prolate model can reproduce the orbital pole precession of the Sgr debris but an oblate potential could. \citet[hereafter LJM05]{Law;et-al_2005} performed simulations using data of the Sgr debris from the 2-Micron All Sky Survey (2MASS) catalogue and found that the kinematics of leading debris was best fit by prolate halos, whereas the trailing debris typically favoured oblate halos. \citet{Prior;et-al_2009a} reached a similar conclusion.
	
\citet{Belokurov;et-al_2006} found an apparent bifurcation within the Sgr debris which \citet{Fellhauer;et-al_2006} argued can only result from a dark halo having a near spherical shape. \citet{Law;et-al_2009} introduced a tri-axial model with a varying flattening profile $q$, which replicates the orbital precession seen and matches kinematic observations of the Sgr debris. However \citet{Law;Majewski_2010} (hereafter LM10) concede this may be a purely numerical solution as tri-axial halos are dynamically unstable. They emphasize that  more kinematic measurements in other regions of the Sgr stream are required.

After the Sgr stellar stream, the Virgo Over-Density (VOD) is arguably the next most significant substructure within our Galaxy. The first over-density in the vicinity of the VOD was observed as a group of RR Lyrae stars by the QUEST survey \citep{Vivas;et-al_2001}. The collaboration later named this the ``$12\fh4$ clump'' \citep{Zinn;et-al_2004}. The broad nature of the VOD was later uncovered from the SDSS catalogue as a diffuse over-density of main-sequence turnoff stars centered at $r_\odot \sim18$ kpc \citep[which ][dubbed as S297+63-20.5]{Newberg;et-al_2002}. The nomenclature on the substructure names within this region is varied, however in this paper when referring to the VOD we are discussing the spatial over-density of stars within the region, separate from any detected co-moving groups.
	
The difficulty arises in accurately distinguishing the VOD as there are multiple substructures along this line-of-sight. \citet{Duffau;et-al_2006} took observations of BHB and RR Lyrae within the ``$12\fh4$ clump'' and found a common velocity of $V_{GSR} = 99.8$ km s$^{-1}$ with $\sigma_{v} = 17.3$ km s$^{-1}$. It should be noted that the dispersion in kinematics measured by \citet{Duffau;et-al_2006} is essentially their velocity precision, so the substructure may possess a much smaller kinematic dispersion. This co-moving group was coined the Virgo Stellar Stream (VSS) to differentiate it from the broad spatial over-density. This distinction from the VOD was somewhat strengthened with new distance measurements which placed the VOD centroid at $r_\odot = 16$ kpc \citep{Juric;et-al_2008, Keller_2010}, and the VSS 3 kpc further away \citep{Duffau;et-al_2006}. Although, when considered in the light of systematic and observational uncertainties, this is of marginal significance. Additionally, \citet{Juric;et-al_2008} suggests the VOD may extend between $r_\odot = 6$ to 20 kpc, which further complicates the matter of distance separation.
 
The relationship between the VSS and the S297+63-20.5 over-density is still unclear. \citet{Newberg;et-al_2007} found a kinematic signature at $V_{GSR} = 130 \pm 10$ km s$^{-1}$ for members of the VOD/S297+63-20.5, which is extremely close to the VSS peak. The VSS and S297+63-20.5 are co-incident in space, but their velocity difference has not yet been reconciled. The distance measurements between S297+63-20.5 and the VSS are similar enough ($\sim1$ kpc) within probable distance uncertainties for \citet{Newberg;et-al_2007} and \citet{Prior;et-al_2009a} to infer they are part of the same structure. \citet{Newberg;et-al_2007} estimate a distance to S297+63-20.5 of $r_\odot = 18$ kpc from $g_0 = 20.5$ turnoff stars, but they concede the structure is likely dispersed along the line-of-sight as the Color-Magnitude Diagram (CMD) for this region does not demonstrate a tight sequence.  Certainly this region of sky, aptly coined the 'Field of Streams' by \citet{Belokurov;et-al_2006}, is complex territory.
	
Photometric studies are inadequate to fully untangle this region. Kinematics are essential to identify co-moving groups that are distinct from the general halo field. Chemical information is vital to accurately distinguish these substructures and understand their origins. However, very few studies have directly investigated metallicities for these stars. In this paper we present spectroscopic observations of K-giants in this region.  Sgr stream kinematics are used to probe the shape of the dark matter halo in the Galaxy. We report both the velocities and metallicities of these giants in an effort to help untangle this accretion-dominated region.

Target selection methodology is outlined in the next section which is followed with details regarding the observations. Techniques used to separate K-giants from dwarfs are discussed in \S\ref{sec:dwarf-giant-separation}, and our analysis procedure for kinematics (\S\ref{sec:kinematics}) and metallicities (\S\ref{sec:metallicities}) follows. A discussion of substructures is outlined in \S\ref{sec:discussion}, and in \S\ref{sec:carbon-stars} we report the carbon stars discovered in our sample. In \S\ref{sec:conclusions} we conclude with some final remarks and critical interpretations. 
		
\section{Target Selection}
\label{sec:target-selection}
	
\begin{figure}[h]
	\includegraphics[width=\columnwidth]{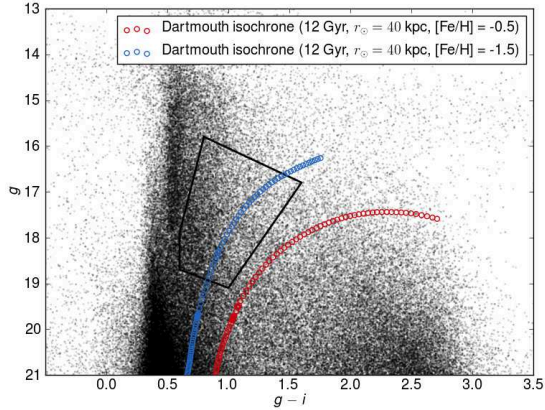}
	\caption{CMD of our observed regions from SDSS DR7, overlaid with the color selection criterion used to target K-giants. Appropriate Dartmouth isochrones \citep{Dotter;et-al_2008} are shown for Sgr debris at a distance of $r_\odot =40$ kpc \citep{Belokurov;et-al_2006}}
	\label{fig:cmd-target-selection}
\end{figure}

When the presence of a stellar substructure is uncovered, K-giants provide excellent candidates for spectroscopic follow-up. They allow for precise radial velocities and chemical abundances. In order to specifically target K-giants we have chosen candidates within the color selection box shown in Figure \ref{fig:cmd-target-selection}, taken from the SDSS DR7 catalogue. This selection box favors a metal-poor population at Sgr stream-like distances ($\sim40$ kpc) but will also contain similarly metal-poor stars at VOD/VSS distances ($\sim20$ kpc) without being significantly biased. Field dwarfs are expected to contaminate the sample due to their similarity in colors. Although K-dwarfs are difficult to distinguish photometrically we can spectroscopically separate these through the equivalent width (EW) of the  gravity-sensitive Mg I triplet lines at 5167.3, 5172.7, and 5183.6 \AA\ (see \S\ref{sec:dwarf-giant-separation}).

\section{Observations}
\label{sec:observations}

Our targets were observed over two runs using AAOmega on the 3.9-m Anglo-Australian Telescope at Siding Springs Observatory in New South Wales, Australia. AAOmega is a double-beam, multi-object (392) fibre-fed spectrograph covering a two degree field of view. The targets were observed in normal visitor mode in April 2009. Throughout all observations, sufficient sky fibres ($\sim30$) were allocated to ensure optimal sky subtraction. In total 3,453 science targets were observed across 4 fields within the VOD/Sgr region, as outlined in Table \ref{tab:observations}. Multiple configurations were observed for most fields to permit measurements of bright ($i < 16$) and faint ($i > 16$) stars, as well as repeat observations on a subset of stars.


\begin{deluxetable}{cccc}
\tablecolumns{1}
\tablewidth{0pt}
\tabletypesize{\scriptsize}
\tablecaption{Observed fields\label{tab:observations}}
\tablehead{
	\colhead{Field} &
	\colhead{$\alpha$-center} &
	\colhead{$\delta$-center} &
	\colhead{Field} \\ 
	& (J2000.0) & (J2000.0) & Configurations
}
\startdata
A & 12 00 00 & $+00$ 00 00 & 1 \\ 
B & 12 20 00 & $-01$ 00 00 & 2 \\ 
C & 12 40 00 & $-02$ 00 00 & 3 \\ 
D & 12 56 00 & $-02$ 42 00 & 6 \\ 
\enddata
\end{deluxetable}

The beam was split into the red and blue arms using the 5700 \AA{} dichroic. The 580V grating in the blue arm yields spectra between 370-580 nm, with a resolution of $R = 1300$. In the red arm we used the 1000I grating ($R = 4400$) which spans the spectral range from 800-950 nm. This coverage includes the Ca II NIR Triplet (CaT), which is used for radial velocities and metallicities. Science targets on each configuration were limited to 1.5 magnitudes in range to minimise scattered-light cross talk between fibres. Globular clusters NGC 5024, 5053 and 5904 were observed as radial velocity and metallicity standards.

The data was reduced using the 2\textsc{DFDR}\footnote{http://www.aao.gov.au/AAO/2df/aaomega} pipeline. After being flat-fielded, the fibres were throughput calibrated and the sky spectrum was subtracted using the median flux of the dedicated sky fibres. Wavelength calibration was achieved from arc lamp exposures taken between each set of science fields. Multiple object frames were combined to assist with cosmic ray removal. 

\section{Dwarf / Giant separation}
\label{sec:dwarf-giant-separation}

	When discussing our data with respect to stellar streams and substructures within the halo, we are referring only to K-type giants.  Dwarfs that fall within our apparent magnitude limit are not sufficiently distant to probe halo substructures. Our resolution is adequate such that the Mg I triplet lines can be individually measured and used to discriminate against dwarfs. 	

A grid of synthetic spectra has been generated to quantitively establish a suitable giant/dwarf separation criterion. The grid was generated using \citet{Castelli;Kurucz_2004} model atmospheres with MOOG\footnote{http://www.as.utexas.edu/$\sim$chris/moog.html} and the line list of \citet{Kurucz;Bell_1995}. Spectra were also generated using stellar parameters for the Sun and Arcturus. The strength of the Mg I lines were tuned to match both the Solar and Arcturus atlases of \citet{Hinkle;et-al_2003}. \citet{Girardi;et-al_2004} isochrones have been used to translate our de-reddened $g - i$ color range to effective temperature. Reddening is accounted for using the \citet{Schlafly;Finkbeiner_2011} corrected dust maps of \citet{Schlegel;Finkbeiner;Davis_1998} assuming a \citet{Fitzpatrick_1999} dust profile where $R_V = 3.1$. The corresponding effective temperature region ranges from 3900 to 5200 K, and is stepped at 25 K intervals. We have assumed typical K-type surface gravities of $\log{g} = 2$ for giants and $\log{g} = 4.5$ for dwarfs. Metallicities of [Fe/H] $= -0.5$, $-1.5$, and $-2.5$ were considered for both surface gravities.

\begin{figure}[h]
	\includegraphics[width=\columnwidth]{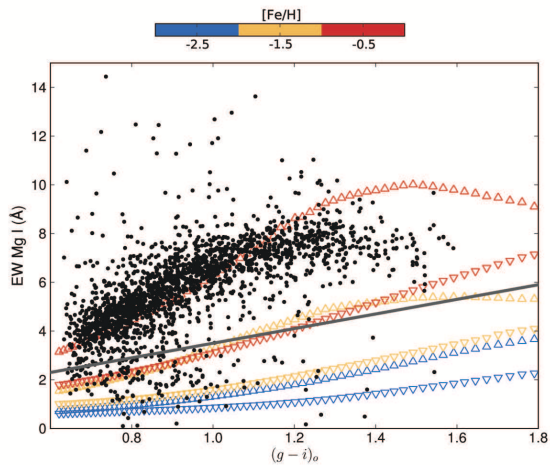}
	\caption{The sum of the Mg I triplet EWs for observations and synthetic spectra shown against the Sloan $g\,-\,i$ color. Synthetic spectra are shown for $\log{g} = 2$ ($\bigtriangledown$) and $\log{g} = 4.5$ ($\bigtriangleup$), and points are colored by metallicity. Our dwarf/giant separation line is shown (solid). }
	\label{fig:dwarf-giant-separation}
\end{figure}

All synthetic spectra was mapped onto the same wavelength intervals before the flux was convolved with a Gaussian kernel of 3.03 \AA{} to match our 580V observations. Mg I line strengths for our observations and synthetic spectra are shown against $g - i$ in Figure \ref{fig:dwarf-giant-separation}. As expected, there is an overlap of Mg I strengths between metal-rich ``giants'' ($\log{g} = 2$) and metal-poor ``dwarfs'' ($\log{g} = 4.5$). However we do not expect metal-poor dwarfs to be a principle contaminant due to their intrinsically low luminosities and comparative rarity.

 Using our synthetic grid, we have varied the slope and offset of our separation line to assess the effectiveness in differentiating giants from dwarfs.  A linear rule has been adopted to differentiate the populations and maximise giant recoverability with minimal contamination. Using the rule,
\begin{equation}
	EW_{Mg\,I} < 3(g-i)_o + 0.5
\end{equation}
\noindent we identify 185 giants in our observed fields. An analysis of our 185 giant candidates revealed three stars with high proper motions \citep[PPMXL Catalog][]{Roser;et-al_2010}, all of which lay close to our dwarf/giant separation line. This suggests they are dwarfs, and they have been excluded. Some possible giant targets were also discarded due to insufficient S/N, or because the Mg I triplet fell on bad columns of the detector. All efforts were made to minimise these losses. The distilled giant sample size is 178 stars.  All identified K-giant stars in our distilled sample had no proper motion detectable above measurement errors.

\section{Radial Velocities}
\label{sec:kinematics}

The Ca II triplet absorption lines at 8498.0, 8542.1 and 8662.1 \AA{}  have been used to measure radial velocities. These lines are strong, and easily identifiable in Red Giant Branch (RGB) stars even at low resolution.  Observed spectra have been cross-correlated with a synthetic spectrum of a typical K-giant ($T_{eff} = 4500$ K, $\log{g} = 2$, [Fe/H] $= -1.5$) to measure radial velocities. Heliocentric corrections were also applied. Radial velocity measurements made of the standard stars in our globular clusters agree (within 3 km s$^{-1}$) with the catalogue of \citet{Harris_1996} (2011 edition). Further, a number of our targets were observed on multiple fields, which allows us to calculate the internal measurement error. The differences between multiple measurements of the same target were calculated and they form a half-normal distribution with a HWHM $= 3.58$ km s$^{-1}$. 
	
In order to compare our kinematic results in a homogenous manner, we have translated our heliocentric velocities to a galactocentric frame. We have adopted the circular velocity of the Local Standard of Rest (LSR) at the Sun as 220 km s$^{-1}$ \citep{Kerr;Lynden-Bell_1986} and accounted for the Sun's peculiar velocity to the LSR by using 16.5 km s$^{-1}$ towards $l = 53^\circ, b = 25^\circ$ \citep{Mihalas;Binney_1981}. The corrected line-of-sight velocity is then given by,
\begin{eqnarray}
	&V_{GSR} = & V_{OBS} + 220\sin{l}\cos{b} + 16.5  \\
	& 		 &\times[\sin{b}\sin{25} + \cos{b}\cos{25}\cos{(l - 53)}] \nonumber
\end{eqnarray}
\noindent where $V_{OBS}$ is the heliocentric-corrected observed line-of-sight velocity. A caveat to this reference transformation is that other authors in the literature have used slightly different formulae to transpose their kinematics to a galactocentric frame. This will result in possible systematic shifts in velocities between authors of up to $\sim11$ km s$^{-1}$. 

\section{Metallicities}
\label{sec:metallicities}

We have measured metallicities for our giants using the strength of the Ca triplet lines. This technique was originally empirically described for individual stars in globular  clusters \citep{Armandroff;Da-Costa_1991}. A spectroscopic analysis using VLT/FLAMES observations of RGB stars from composite populations led \citet{Battaglia;et-al_2008} to conclude that a calibrated CaT-[Fe/H] relationship can be confidently used in composite stellar populations \citep[see also][]{Rutledge;Hesser;Stetson_1997, Starkenburg;et-al_2010}. The caveat to this technique is that a luminosity (specifically $V - V_{HB}$) is required for calibration, and we have to assume a $V_{HB}$ luminosity here. Johnson V-band magnitudes for our giants were calculated from SDSS \textit{ugriz} photometry using \citet{Jester;et-al_2005} transformations. The weaker third Ca triplet line is more susceptible to noise and residual sky-line contamination \citep{Tolstoy;et-al_2001,Battaglia;et-al_2008}. Consequently, only the strongest two CaT lines (8542 and 8662\AA) have been used to form a reduced EW (W') such that,

\begin{eqnarray}
	\textstyle\sum{W}\, &=& EW_{8542} + EW_{8662} \\
	W' &=&\textstyle\sum{W} + 0.64\left(\pm 0.02\right)\left(V-V_{HB}\right)
\end{eqnarray}
\noindent and the metallicity linearly varies with W' where,
\begin{equation}
	\mbox{[Fe/H]}_{\mbox{\small{CaT}}} = (-2.81\pm0.16) + (0.44\pm0.04) W'
	\label{eq:feh-cat}
\end{equation}

Using this calibration our globular cluster standard stars (with known $V_{HB}$ magnitudes) have metallicities that match well with the \citet{Harris_1996} catalogue (2011 edition). Only K-giants within the valid calibration range ($0 > V-V_{HB} > -3$) were considered for metallicities. We assume that we have two dominant substructures present in our observations; the leading arm of Sgr and the Virgo Over-Density (see \S\ref{sec:substructure-identification}). The Sgr stream dominates our negative $V_{GSR}$ population, and our positive kinematic space primarily comprises VOD and VSS members. As such we have separated our population at $V_{GSR} = 0$ km s$^{-1}$ into two samples with assumed $V_{HB}$ magnitudes. Although this introduces a (known) systemic effect, it is a required assumption to estimate the metallicity distribution of each population.

The horizontal branch magnitude assumed for the VSS/VOD has been ascertained from previous RR Lyrae studies. As RR Lyrae stars sit on the horizontal branch, the $V_{HB}$ is taken as the V-band median of four \citet{Prior;et-al_2009a} and three  \citet{Duffau;et-al_2006} VSS members to yield $\langle{}V\rangle{} = 17.09$. This combined value precisely matches the mean reported by \citet{Duffau;et-al_2006}. In our K-giant sample with positive galactocentric velocities, we assume $V_{HB} = 17.09$ which implies these stars are at a distance of $\sim20$ kpc.

As the distance to the Sgr debris varies greatly throughout the halo, the horizontal branch magnitude varies with the position along the stream. 
Through CMD comparisons with the Sgr core \citep{Bellazzini;et-al_2006}, \citet{Belokurov;et-al_2006} determined distances along the two branches of the Sgr bifurcation.  Revised distance measurements by \citet{Siegel;et-al_2007} (as tabulated in Table 1 of LM10) have been crucial in constraining dark halo models. Our observations lay on the edge of Branch A (Figure \ref{fig:sgr-field-of-streams}).

We have assumed distances to these stars by interpolating their position along a cubic spline fitted to published distances \citep{Siegel;et-al_2007}. Recent distance measurements of the stream by \citet{Ruhland;et-al_2011} match the stream distances used here within the uncertainties. The horizontal branch magnitude $V_{HB}$ is calculated using this assumed distance and the known luminosity of RR Lyrae stars \citep[$M_V = +0.69;$][]{Tsujimoto;et-al_1998}. Typical $V_{HB}$ magnitude derived for the Sgr members is $18.7$, corresponding to a distance of $\sim{}41$ kpc. Reddening is accounted for using the \citet{Schlegel;Finkbeiner;Davis_1998} maps with \citet{Schlafly;Finkbeiner_2011} corrections as described in \S\ref{sec:dwarf-giant-separation}. Uncertainties in distance are not interpolated; they are taken as the largest published uncertainty of the closest neighbouring data points, and this is propagated through to our reported metallicity uncertainties. The kinematics and metallicities derived for our giant sample are discussed in the following section.

\section{Discussion}
\label{sec:discussion}
		
	\subsection{Substructure Identification}
	\label{sec:substructure-identification}
		
\begin{figure}[h!]
		\includegraphics[width=\columnwidth]{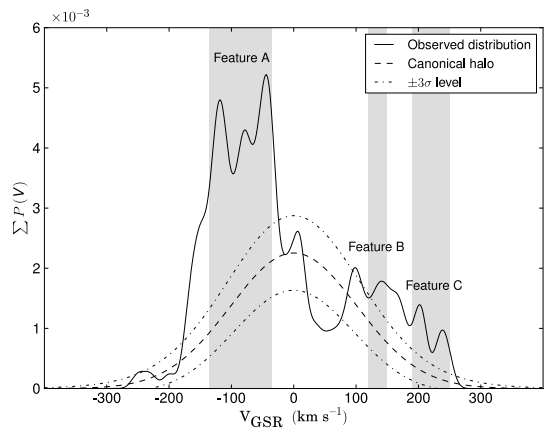}
		\caption{Generalised histogram of $V_{GSR}$ for our 178 K-type giants, highlighting the substructure present. The Gaussian represents the halo contribution; it is normalised such that the integral equals the number of observed stars excluding those outside a 2.5-$\sigma$ excess. Significant ($>3\sigma$) kinematic deviations from the smooth distribution are appropriately grouped and labelled.}
		\label{fig:velocity-histogram}
	\end{figure}

A significant kinematic deviation from a canonical halo population signifies a co-moving group. There are multiple substructures along our line-of-sight. These features are identified first and discussed separately. We have represented our galactocentric velocities with a generalised histogram (Figure \ref{fig:velocity-histogram}) to quantitatively compare the observed stellar kinematics against the hypothesis that the parent distribution can be described by a canonical halo where $\mu = 0$ km s$^{-1}$, and $\sigma = 101.6$ km s$^{-1}$ \citep{Sirko;et-al_2004}.

		\begin{figure*}[t!]
		\includegraphics[width=\textwidth]{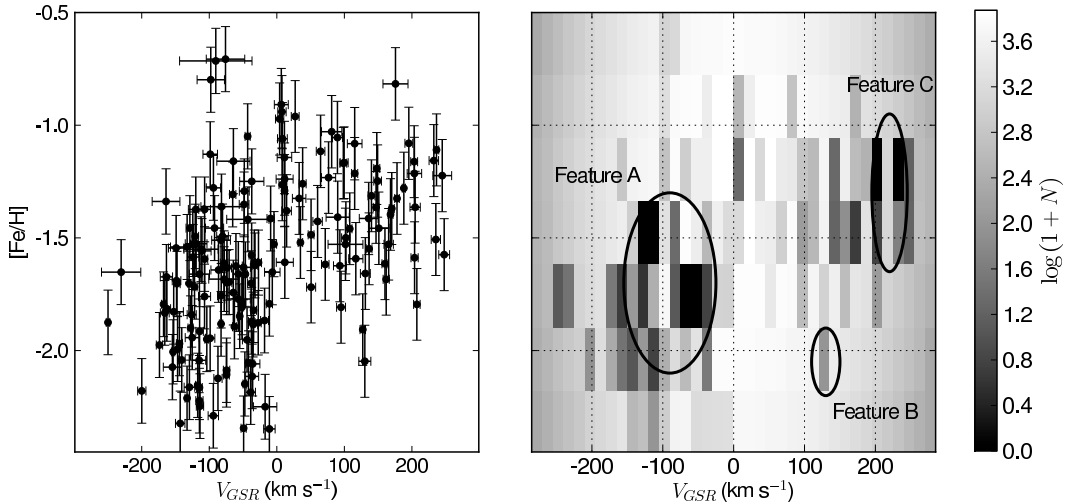}
		\caption{The observed data for the 178 giant star sample (left panel). Monte-Carlo simulation results illustrating the number of times simulations could reproduce our observed data in the equivalent multi-dimensional bin (right panel). Significant features discussed in the text are labelled.}
		\label{fig:monte-carlo}
	\end{figure*}

The generalised histogram represents each data point with a Gaussian kernel of an equal deviation. As our internal kinematic errors between multiple measurements summate to a half-normal distribution with a FWHM of 7.16 km s$^{-1}$, we have opted for a sigma value of 10 km s$^{-1}$ for the generalised histogram in Figure \ref{fig:velocity-histogram} to avoid under-smoothing. The most significant ($>3\sigma$) kinematic peaks have been labeled Features A, B, and C. Features A and C are our most significant structures, and the importance of Feature B becomes more evident through Monte-Carlo simulations.

	Velocities and metallicities of our K-giants, shown in the left panel of Figure \ref{fig:monte-carlo}, were binned into grid blocks of 15 km s$^{-1}$ and 0.3 dex \--- roughly twice the error in each dimension. To interpret these data a population of 178 stars was randomly drawn from a simulated halo with canonical kinematics. Metallicities were randomly assigned using the observed halo Metallicity Distribution Function (MDF) of \citet{Ryan;Norris_1991}. Each simulated population was binned identically to our observed sample. Simulation grid blocks with counts equal to or exceeding stars in the equivalent observed grid block were noted, and summed after 10,000 simulations. Results from our Monte-Carlo simulations are illustrated in Figure \ref{fig:monte-carlo} (right). The identified Features A, B, and C in our results were consistently significant when the grid was midpoint offset in both dimensions, except for the grid-block centered at $\sim185$ km s$^{-1}$ and $\sim-1.4$ dex, which has consequently been left unlabelled.
	
	 Substructure becomes statistically significant when the observed grid blocks are rarely replicated in Monte-Carlo simulations. Specifically, the number of members in Feature A was never reproduced in some grid bins. This feature is well in excess of the halo and has a wide spread in kinematics. \citet{Chou;et-al_2007} mapped the Sgr debris across the sky using K/M-giants and found galactocentric velocity signatures between $-205$ km s$^{-1}$ to $-31$ km s$^{-1}$ in this region. As such we attribute this wide, significant kinematic peak as the leading arm of the Sgr tidal tail. Feature A is discussed in the next section. Another feature where grid blocks were never reproduced in our simulations was Feature C. This feature may also be attributed to Sgr debris and is discussed further in \S\ref{sec:feature-c}.
	 	
	 The clump of stars in the velocity bin centered on $V_{GSR} \approx 130$ km s$^{-1}$ and [Fe/H] $\approx -2$ was replicated only $\sim1$\% in 10,000 simulations. We attribute these stars to the Virgo Stellar Stream, as they are coincident in spatial position, velocity, and metallicity with previously reported values of the VSS obtained by F turnoff/BHB stars \citep{Newberg;et-al_2007} and RR Lyrae stars \citep{Prior;et-al_2009a}. Feature B is discussed in \S\ref{sec:the-vss}.

	\begin{figure*}[t!]
		\includegraphics[width=\textwidth]{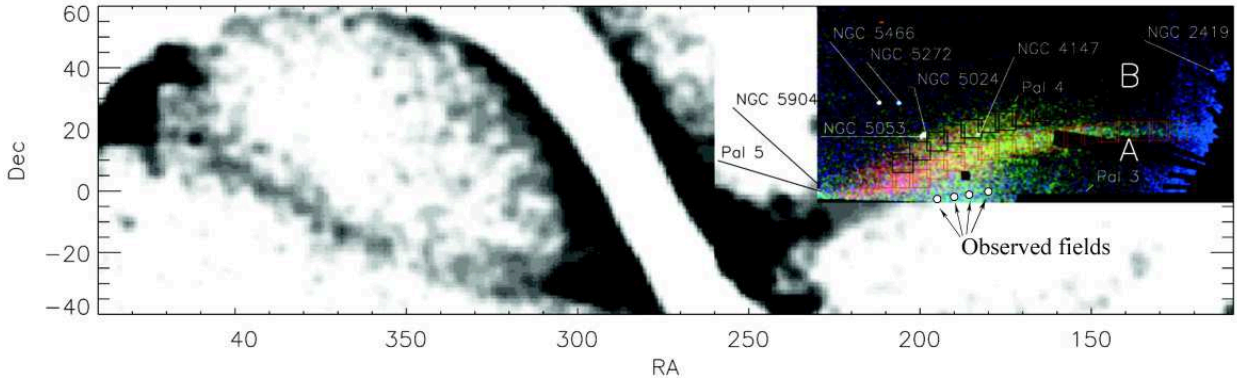}
		\caption{Observed fields are outlined upon a panoramic view of the Sgr stream, to demonstrate our field locations in context with the Sgr stream. This plot is an adaptation of Figure 2 in \citet{Belokurov;et-al_2006}, which uses the 2MASS M-giant sample of \citet{Majewski;et-al_2003}.}
		\label{fig:sgr-field-of-streams}
	\end{figure*}
	
	\subsection{Feature A \--- Sagittarius Debris}
	\label{sec:sgr-debris}
	
	  Our observed fields are overlaid upon a panoramic projection of the Sgr stream and the  ``Field of Streams''  \citep{Belokurov;et-al_2006} in Figure \ref{fig:sgr-field-of-streams}. This is a crowded region of globular clusters, substructures and overlapping stellar streams, primarily populated by the Sgr Northern leading arm. Simply from a spatial perspective, we expect the Sgr debris to dominate our data. Although the VOD is present, it is much more diffuse. 
	
	\subsubsection{Comparing Sagittarius Debris to Dark Matter Halo Models}	

	In order to investigate the spatial coverage and kinematics of our Sgr members we have compared our data with the constant flattening spheroidal models of LJM05, and the more recent tri-axial model of LM10. The simulated data output from these models is readily available online\footnote{http://www.astro.virginia.edu/$\sim$srm4n/Sgr/}, and the released models have the best-fitting parameters for each dark halo shape (prolate, spherical, oblate and tri-axial).  These are the only simulations available which make use of an all-sky data set; the 2MASS M-giant sample. We have not considered particles farther than 60 kpc to match realistic observation limitations.  The $10^5$ model Sgr debris particles from their simulations are represented along the best-fit great-circle ($\Lambda_\odot$, $B_\odot$) where the longitudinal coordinate $\Lambda_\odot$ is zero in the direction of the Sgr core and increases along the trailing debris. For comprehensive details of the simulations the reader is referred to the papers of \citet{Law;et-al_2005} and \citet{Law;Majewski_2010}. 
		
	\begin{figure}[t!]
		\includegraphics[width=\columnwidth]{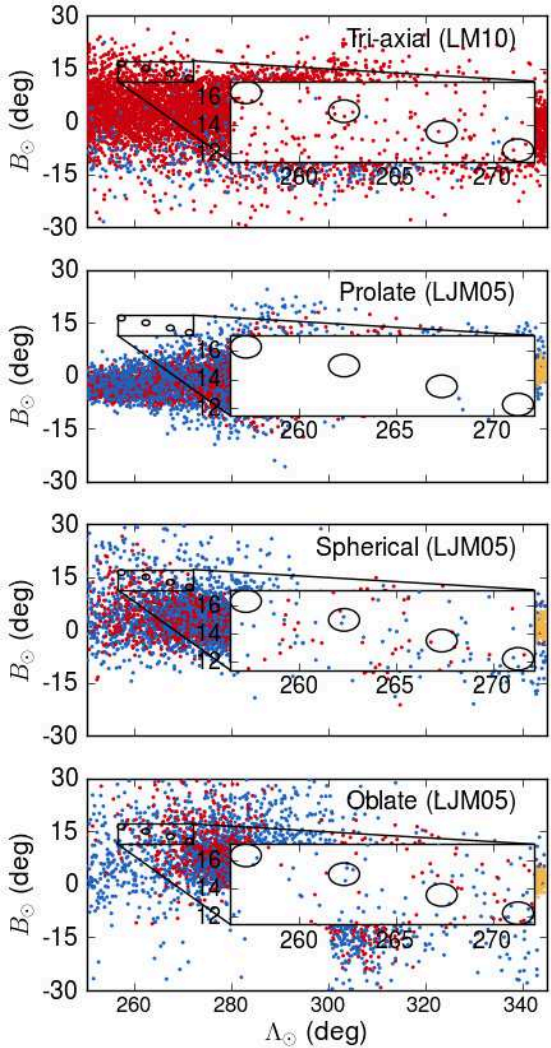}
		\caption{Our fields and observed bounds are shown with the simulated particles from multiple \textit{N}-body simulations by LJM05 and LM10. These particles are distance restricted (up to 60 kpc), and color-coded by their peri-centric passage (yellow; most recent passage, red; previous passage, blue; oldest observable passage).}
		\label{fig:law-spatial}
	\end{figure}

	The Northern leading arm of Sgr is particularly sensitive both kinematically and spatially to the shape of the Milky Way dark halo (Figure \ref{fig:law-spatial}). The spherical, oblate and tri-axial models predict the Sgr stream to pass directly through our fields, whereas the prolate model predicts only the edge of the stream near this region and no particles directly in our fields. Spatially, the prolate model predicts the Sgr stream to pass much lower in latitude ($B_\odot$) than our observed fields. When we extend a rectangle bounded by the edges of our fields (as per the zoomed insets in Figure \ref{fig:law-spatial}) a mere two model particles are found. The $10^5$ model particles are proportional to the stream density so the lack of model particles implies a negligible stream concentration in our observed region. Observationally, when the number of field configurations is accounted for our observed Sgr K-giants are uniformly distributed across the fields. If this prolate halo model is a true representation of the Sgr debris, this does not exclude the potential of finding some Sgr debris in our fields. However, it does imply that if the LJM05 prolate model is an accurate representation of the dark halo then Sgr would not be the dominant population \--- contrary to our preferred interpretation of the observations.
	
	In comparison, the tri-axial, spherical and oblate models predict varying amounts of debris from previous peri-centric passages in our fields. A kinematic comparison against our observations is necessary to evaluate model predictions. As the prolate model has only two simulated particles within our extended bounds there is no qualitative kinematic comparison to be made for this model, therefore the prolate model is excluded from further kinematic analysis.

	\begin{figure*}[t!]
		\includegraphics[width=\textwidth]{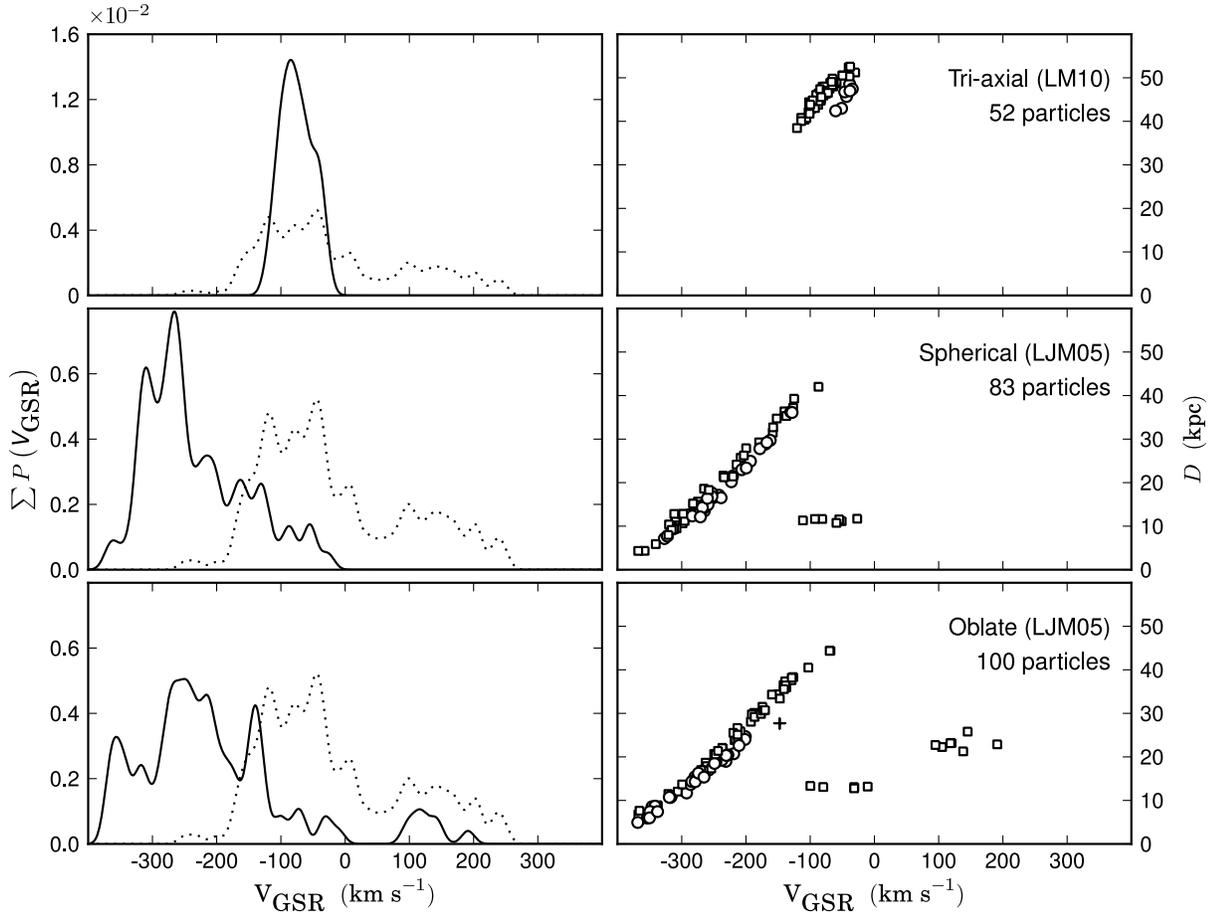}
		\caption{A generalised histogram (left) of the galactocentric rest frame velocities in our sample (dotted line) compared to the velocities (solid line) from the \textit{N}-body models of LJM05, LM10 of particles within the same spatial coverage and distance range (up to $r_\odot \sim 60$ kpc) of our observed sample. Heliocentric distances for particles used to generate each velocity histogram are shown on the right. Particles are marked by their peri-centric passage. A plus ($+$) denotes debris from the current peri-centric passage by Sgr, circles ($\circ$) mark the previous passage, and squares ($\square$) represent debris from two previous passages. The prolate model from LJM05 is not considered in this plot (see text).}
		\label{fig:law-vel-compare}
	\end{figure*}
	
	The number of predicted particles observable differs between models. Consequently, the kinematics for each model have been represented as a generalised histogram (Figure \ref{fig:law-vel-compare}). For consistency all simulation output and observed data has been convolved with a Gaussian kernel of 10 km s$^{-1}$; the sigma used for the previous generalised histogram in Figure \ref{fig:velocity-histogram}. Predicted distances and peri-centric ages are shown, demonstrating that debris from multiple passages are predicted along the line-of-sight. A much tighter velocity sequence is predicted by the tri-axial model than the constant flattening models. This is because the tri-axial model predicts a peri-center almost precisely in our observable region, whereas the spherical and oblate models predict a large wrap along the line-of-sight, resulting in a wide spread of distances and velocities.
	
	Predicted distances are consistent with what we would expect from our observations. The median luminosities of our giants is $g\sim17.5$, so for the tri-axial model particles at $45$ kpc this corresponds to an absolute magnitude of $M_g\sim-0.8$; quite reasonable for a K-giant. Similarly for the spherical and oblate models, typical distances of $\sim25$ kpc will yield luminosities of $M_g\sim+0.5$ which is also reasonable.  
	
	The spherical and oblate models also predict close-by debris with extremely negative galactocentric velocities. This signature is not represented in our data. The lowest observed velocity is $\sim{}-250$ km s$^{-1}$ and we have only two observations less than $V_{GSR} < -200$ km s$^{-1}$. In contrast, the spherical/oblate models predict velocities well below $V_{GSR} < -300$ km s$^{-1}$. These predicted highly negative  velocities were also not found in Sgr RR Lyrae observations taken by \citet{Prior;et-al_2009b}. If the dark matter potential is well-represented by either an oblate or spherical halo, then this discrepancy must be reconciled. 
	
	The oblate and spherical model also illustrate a similar signature at $r_\odot \sim$12 kpc, with particle velocities ranging between $-100 < V_{GSR} < 0$ km s$^{-1}$. This is the edge of a predicted crossing-point between different wraps of the stream, which occurs at $\sim12$ kpc. These particles (and the positive kinematic signature around $\sim$20 kpc in the oblate model), are relatively minor signatures when compared to the Northern arm feature. If these signatures are present in our observed fields, their relatively low density compared to the Northern arm feature would prevent them from appearing as significant. 
	
	Although the predicted velocity distribution is narrower than what we observe, the LM10 tri-axial dark halo model best fits our observations. The observed sample broadens most prominently towards more negative galactocentric velocities at $\sim-200$ km s$^{-1}$, considerably higher than oblate/spherical model predictions. Halo contamination at highly negative velocities is likely to be small. One reconciliation for this discrepancy between our observations and the tri-axial model may lie in the workings of the tri-axial model itself. Unlike other models considered, the tri-axial model does not reproduce the observed bifurcation in the Sgr stream \citep{Belokurov;et-al_2006}. Although observationally untested, it is reasonable to suggest a bifurcation may result from a significant kinematic disruption. Such an effect could result in a broader kinematic distribution \--- similar to what we have observed \--- which is most notable at the stream edges.
	
	There is further work required through observations and simulations to reconcile kinematic discrepancies. Typical examinations of the leading arm debris usually favour prolate halos and evidence along the trailing arm typically favours oblate halos \citep{Helmi_2004, Martinez-Delgado;et-al_2004, Law;et-al_2005}. Kinematic predictions of the LM10 tri-axial model reasonably match our observations, whereas the prolate model has been excluded as significant debris is not predicted in this edge of the stream. These observations along the Northern leading arm are the first which are not reproducible with the current prolate model of LJM05, contrary to previous groups who have surveyed closer to the leading arm debris.

	\subsubsection{A Metal-Poor Population Uncovered in Sagittarius Debris}
	\label{sec:sgr-metal-poor}
		
	Many groups have found the population of the Sgr core to possess a mean [Fe/H] $\sim -0.5$ dex \citep{Cacciari;et-al_2002, Bonifacio;et-al_2004, Monaco;et-al_2005}. The observed metallicity gradient along the stream suggests that as the host circles the Milky Way the older, more metal-poor stars are preferentially stripped from the progenitor \citep{Chou;et-al_2007, Keller;Yong;Da_Costa_2010}. In this region of the stream very few Sgr member metallicies have been reported. \citet{Vivas;et-al_2005} found a mean metallicity of $\langle$[Fe/H]$\rangle = -1.77$ from spectra of 16 RR Lyrae stars along a nearby region of the Sgr leading arm. Similarly, \citet{Prior;et-al_2009b} found $\langle$[Fe/H]$\rangle = -1.79 \pm 0.08$ for 21 type \textit{ab} RR Lyrae stars in the region. This is somewhat expected since only the oldest, most metal-poor stars can form RR Lyraes. 
	
\begin{figure}[h]
		\includegraphics[width=\columnwidth]{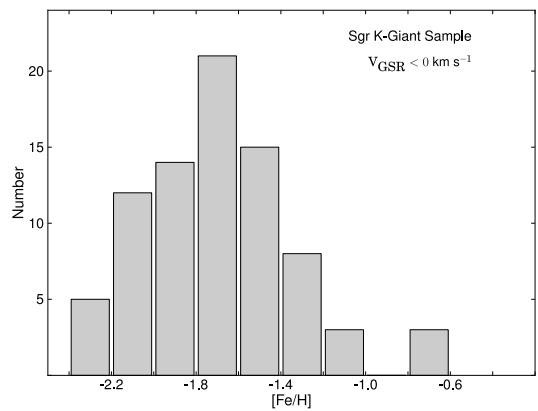}
		\caption{Metallicity histograms for K-giant members in our negative galactocentric velocity sample, which is largely populated by Sgr debris from the Northern leading arm.}
		\label{fig:sgr-metallicity-hist}
	\end{figure}

	Investigating the MDF of the Sgr debris requires an unbiased sample. Generally K-giants are excellent stellar candidates for such investigations as all stars go through this evolutionary phase whereas M-giants are consistently metal-rich. If we apply the metallicity gradient found by \citet[][from M-giants]{Keller;Yong;Da_Costa_2010} to these observed K-giants, we would expect an abundance mean near $\sim{}-1.2$ dex in this observed region. The metallicity distribution for our entire negative $V_{GSR}$ sample is shown in Figure \ref{fig:sgr-metallicity-hist}, and illustrates a metal-poor population. The mean of our distribution is $\langle$[Fe/H]$\rangle = -1.7\,\pm\,0.3$ dex. If we include only the stars attributed to Feature A (as defined by the shaded region in Figure \ref{fig:velocity-histogram}, between $-140 < V_{gsr} < -30$ km s$^{-1}$) this value increases by only 0.04 dex; well within observational uncertainties. As a comparison, in a sample of metal-rich biased M-giants from the 2MASS data \citet{Chou;et-al_2007} found a mean metallicity of $\langle$[Fe/H]$\rangle = -0.72$ dex for their best subsample in the Northern leading arm of Sgr. Although there may be some halo contamination in our sample, in general our more metal-poor Feature A members have higher negative velocities, which is expected for Sgr stream members in this region.
		
	It is likely that our observed K-giant sample is biased towards more metal-poor members. Given a typical distance of $r_\odot\sim 40$ kpc for the Sgr debris in this region \citep{Belokurov;et-al_2006}, a 12 Gyr old Dartmouth \citep{Dotter;et-al_2008} isochrone with a metallicity of [Fe/H] $= -1.5$ falls directly within our target selection window. The more metal-rich isochrone with [Fe/H] $= -0.5$ does not pass through our selection range. This is consistent with our observed metallicities for Sgr. Although this does not prohibit the possibility of more metal-rich members within our sample, it implies that our distribution is biased towards the metal-poor end of the MDF. The tail of our Sgr distribution extends from $-0.71$ to $-2.35$ dex.

	\subsection{Feature B \--- The Virgo Stellar Stream}
	\label{sec:the-vss}
	
	Another substructure which was distinguished from the general over-density of the VOD is the Virgo Stellar Stream \citep{Duffau;et-al_2006}. Several RR Lyrae stars were noted with a common velocity of $V_{GSR} = 130$ km s$^{-1}$ \citep{Newberg;et-al_2007, Prior;et-al_2009a}. \citet{Prior;et-al_2009a} also measured metallicities of stars with this kinematic signature and found a mean [Fe/H] $= -1.95\,\pm\,0.1$ for the VSS, although this is based on RR Lyrae stars and  therefore is potentially biased to the oldest, most metal-poor populations. The RR Lyrae sample from \citet{Duffau;et-al_2006} found a $\langle$[Fe/H]$\rangle = -1.86$ with a much larger abundance spread of $\sigma = 0.40$, which was several times the relative uncertainty in individual values ($\sigma_{\mbox{[Fe/H]}} = 0.08$ dex). This led \citet{Duffau;et-al_2006} to infer the progenitor of the VSS was likely a dSph galaxy, as a large dispersion in $\mbox{[m/H]}$ indicates self-enrichment and implies the structure is a galaxy and not an isolated stellar cluster. Although some globular clusters are also now known to contain multiple populations and sizeable $\mbox{[m/H]}$ dispersions, globular clusters in general  have a much smaller dispersion than that observed here. Similar to \citet{Duffau;et-al_2006}, we  also observe an abundance spread within this kinematic bin including two K-giants with mean kinematics and metallicities ($\langle{}V_{GSR}\rangle{} = 130 \pm 9$ km s$^{-1}, \langle\mbox{[Fe/H]}\rangle = -2.0 \pm 0.16$) matching those found by \citet{Prior;et-al_2009a}. Monte-Carlo simulations reproduced these targets a mere $\sim1$\% in 10,000 simulations \--- highlighting their significance.

	High resolution follow-up spectroscopy on these targets and other VSS candidates at higher metallicities in this kinematic range will provide crucial information about the origin of the VSS.  Investigating [$\alpha$/Fe] ratios in these candidates can constrain the mass of the VSS progenitor and discern whether the host was indeed a dSph \citep{Venn;et-al_2004, Casetti-Dinescu;et-al_2009}.  Future observations are planned.

	\subsection{Feature C \--- Sagittarius debris?}
	\label{sec:feature-c}

	The last significant kinematic substructure we identify in our data is Feature C. There are two peaks identified here at $V_{GSR} = 200$ and $240$ km s$^{-1}$. These features cannot be explained by a smooth halo distribution. \citet{Newberg;et-al_2007} identified a 4-$\sigma$ peak at $V_{GSR} = 200$ km s$^{-1}$ in their sample of F turnoff stars centered on $(l, b) = (288^\circ, 62^\circ)$. Our nearest field to their plate (Field B) hosts only one star associated with this peak; most of our stars are located within our most populated Field D. \citet{Prior;et-al_2009b} also noted stars in this kinematic range and compiled a list of authors who have similarly observed such peaks (see \citet{Sirko;et-al_2004, Duffau;et-al_2006, Starkenburg;et-al_2009}). \citet{Prior;et-al_2009b} argue these stars may be associated with trailing debris of Sgr, as suggested by models (LJM05, LM10).  Although this trailing debris would be much closer than the nearby visible leading arm, our observations support the interpretation by \citet{Prior;et-al_2009b} as the metallicities are quite similar to the nearby Sgr debris.
	
	This substructure has a range of metallicities. These metallicities were derived assuming a single distance modulus to the VOD/VSS. The abundance dispersion we observe is either representative, or these stars have a common metallicity and are dispersed along the line-of-sight. Unfortunately neither can be positively excluded without further observations. If these peaks are associated with the trailing debris of Sgr debris, they are likely to be further away than the VOD. On our scale that would make these stars more metal-rich than shown in Figure \ref{fig:monte-carlo}. However, the nearest trailing debris particles predicted from any best-fitting dark halo potential that could explain this kinematic peak occur at $\sim\Lambda_\odot = 310^\circ$; our observations range in longitude  between $\Lambda_\odot = 256-273^\circ$. Whether this substructure is separate from Sgr debris or not remains equivocal. Nevertheless, the apparent metallicities of the Feature C stars are not inconsistent with those of a Sgr population given our results for Feature A and those of \citet{Keller;Yong;Da_Costa_2010}.
	
\section{Carbon Stars}
\label{sec:carbon-stars}	

	In our sample we have identified five carbon stars. Our colour selection to select K-giants overlaps with where we would expect to find carbon stars. Although these stars were not specifically targeted, their surface densities are so low \citep[$\approx$ 1 per 50 deg$^2$;][]{Green;et-al_1994} that we would not expect them as a substantial contaminant.  Although our sky coverage is small ($\sim$12.5 deg$^2$), the observed carbon star spatial density is $\sim20$x higher than expected. All five carbon stars are recognisable by the presence of distinctive 4737- and 5165 \AA\, Swan C$_2$ bands.

\begin{figure}[h!]
	\includegraphics[width=\columnwidth]{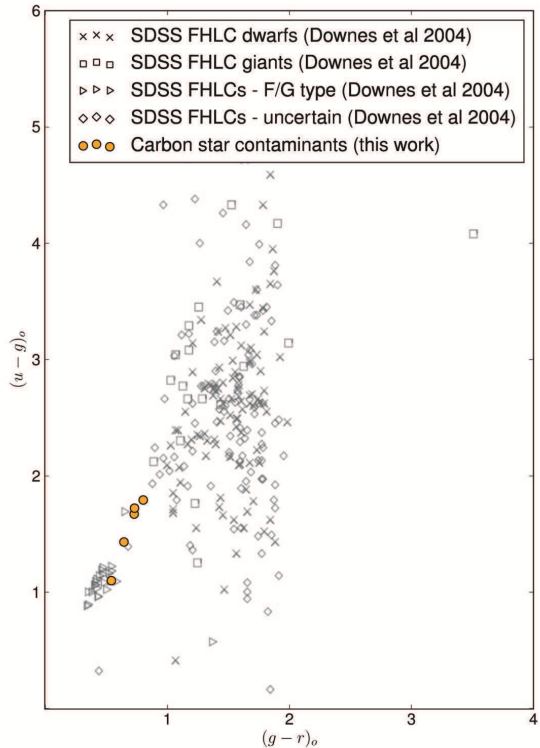}
	\caption{Sloan \textit{u \--- g} and \textit{g \--- r} colors for our carbon stars, and the identified carbon stars and classifications by \citet{Downes;et-al_2004}. Carbon stars identified in this work have been shaded for clarity.}
	\label{fig:carbon-sdss}
\end{figure}

	 Dwarf carbon stars exhibit a spectral signature which mimics that of a typical CH-type giant carbon star, however they have anomalous JHK colors \citep{Green;et-al_1992} and high proper motions. The existence of the 4300 \AA{} CH G-band is representative of a CH-type carbon star, and is found in all five of our carbon stars. SDSS photometry for our stars match well with the F/G-type CH carbon stars identified by \citet{Downes;et-al_2004}, as seen in Figure \ref{fig:carbon-sdss}. However these targets were not spectroscopically observed in the SDSS follow-up survey SEGUE.  Through comparisons with previous carbon-type star catalogues \citep{Totten;Irwin_1998, Downes;et-al_2004, Goswami;et-al_2010}, the stars tabulated in Table \ref{tb:carbon-stars} are previously unclassified carbon stars. This is largely because our objects are too faint to have been observed by previous spectroscopic carbon star surveys.
	
\begin{deluxetable}{cccccccccccc}
\tablewidth{0pt}
\tabletypesize{\scriptsize}
\tablecaption{Properties of carbon stars found in our sample \label{tb:carbon-stars}}
\tablehead{
	\colhead{SDSS Name J+\tablenotemark{a}} & 
	\colhead{$M_g$} &
	\colhead{\it{u-g}} &
	\colhead{\it{g-r}} &
	\colhead{\it{r-i}} &
	\colhead{\it{i-z}} &
	\colhead{\it{H-K}} &
	\colhead{\it{J-H}} &
	\colhead{$\mu_{\alpha\cos{\delta}}\tablenotemark{c}$} &
	\colhead{$\mu_\delta\tablenotemark{c}$} &
	\colhead{$V_{GSR}$} & 
	\colhead{Likely} \\
	& & & & & & & & (mas yr$^{-1}$) & (mas yr$^{-1}$) & (km s$^{-1}$) & dwarf?
} 

\startdata
121740.94-001839.5 & 18.66 & 1.10 & 0.54 & 0.16 & 0.05 & ...\tablenotemark{b}& ...\tablenotemark{b} & \phn$-3.0 \pm 5.4$  & $-18.0 \pm 5.4$\phn & $\phn-71 \pm 34$ & Yes\\
121853.18-004628.4 & 17.71 & 1.67 & 0.73 & 0.22 & 0.11 & \phd0.24 & 0.36 &  \phn$-3.7 \pm 4.5$ & $-0.5 \pm 4.5$ & $+176 \pm 9$ & - \\
122053.71-011709.5 & 17.65 & 1.79 & 0.80 & 0.26 & 0.13 & \phd0.16 & 0.56 &  \phn$-7.3 \pm 4.3$ & $-3.9 \pm 4.3$ & $\phn+31 \pm 11$ & - \\
125410.80-032744.0 & 18.59 & 1.43 & 0.64 & 0.22 & 0.05 & \phd0.56 & 0.32 &$-21.1 \pm 5.0 $ & $-7.7 \pm 5.0$ & $\phn-41 \pm 9\phn$ & Yes \\ 
125416.52-031437.6 & 16.85 & 1.72 & 0.73 & 0.25 & 0.11 & -0.11 & 0.50 & $-19.8 \pm 4.2$ & \phd\phn$8.7 \pm 4.2$& $\phn+15 \pm 4\phn$ & Yes \\
\enddata
\tablenotetext{a}{To keep with convention, positional information has been truncated. Full information available through the SDSS archive.}
\tablenotetext{b}{No \textit{JHK} photometry is available for this object as it is fainter than the 2MASS survey limit.}
\tablenotetext{c}{Proper motions taken from the PPMXL Catalog \citep{Roser;et-al_2010}.}
\end{deluxetable}

	There is 2MASS photometry available for four of these stars. Of those, two stars (SDSS J125410.80-032744.0 and J125416.52-031437.6) exhibit anomalous JHK colors and significant ($> 3\sigma$) proper motions \citep[PPMXL][]{Roser;et-al_2010}; characteristics of a dwarf carbon star. A third star also exhibits significant proper motion (SDSS J121740.94-001839.5), however JHK photometry is unavailable. The two remaining stars in our sample have JHK photometry characteristic of the F/G type CH carbon stars found by \citet{Downes;et-al_2004} in their Faint High-Latitude Carbon (FHLC) star survey (Figure \ref{fig:carbon-sdss}), and do not exhibit significant proper motion - which strongly suggests they are not dwarfs \citep{Green;et-al_1994, Deutsch_1994}.

\section{Conclusions}
\label{sec:conclusions}

	We present spectroscopic observations of K-type stars in a region of the 'Field of Streams', where significant substructure is present. We utilise the gravity-sensitive Mg I triplet to separate giants and dwarfs. Radial velocities and metallicities of 178 K-giants have been determined. We have recovered kinematic substructure found by other authors, which cannot be explained by a smooth halo distribution. Highly negative velocity signatures match those expected by the Sgr stream debris, and we also identify a group of K-giants with metallicities and kinematics which make them highly probable members of the Virgo Stellar Stream.

	Stars in this region of the Sgr stream are kinematically sensitive to the shape of the Galactic dark halo. As such, we have compared our velocity distribution to the Sgr-Milky Way dark matter models of \citet{Law;et-al_2005, Law;Majewski_2010}. Typically, leading arm debris favour prolate models and trailing arm debris favour oblate models. A prolate dark halo predicts relatively little debris in our observed fields. If the spheroid is prolate and LJM05 presents an accurate representation, Sgr debris would not be expected to dominate our sample even if we were $\Delta{}B_\odot\sim10^\circ$ closer to the great circle best-fit. However, Sgr debris is our most significant kinematic structure observed. No single model perfectly represents our data, although we find the more recent tri-axial model (LM10) best represents our observed kinematics. 
	
	Observed metallicities for K-giants in this region of the Sgr stream are notably lower than expected based on other Sgr samples, demonstrating the presence of a metal-poor population in the  Sgr debris. Although isochrones indicate we are biased towards metal-poor members at these distances, these stars are unequivocally Sgr in origin. Metallicity gradients reported, suggest a mean of $\langle[$Fe/H$]\rangle\sim{}-1.2$, whereas our population has $\langle[$Fe/H$]\rangle = -1.7 \pm 0.3$ dex. Previously reported abundances in this region have been ascertained from observing RR Lyrae stars (where the mean metallicity found was $\langle\mbox{[Fe/H]}\rangle\sim{}-1.7$ dex) or from observing M-giants (where $\langle\mbox{[Fe/H]}\rangle\sim{}-0.7$ dex was reported). In both scenarios the measurements are reasonable with what we would expect from each particular stellar population. Thus the sample presented here is more representative of the complete MDF, and less susceptible  to age or metallicity biases. These observations also demonstrate that the Sgr stream likely hosts a substantially larger abundance range than previously found, including a previously un-detected metal-poor population of K-giants.

\section{Acknowledgements}
ARC acknowledges the financial support through the Australian Research Council Laureate Fellowship 0992131. SK and GDaC acknowledge the financial support from the Australian Research Council through Discovery Program DP0878137.

This publication makes use of data products from the Two Micron All Sky Survey, which is a joint project of the University of Massachusetts and the Infrared Processing and Analysis Center/California Institute of Technology, funded by the National Aeronautics and Space Administration and the National Science Foundation.

Funding for the SDSS and SDSS-II has been provided by the Alfred P. Sloan Foundation, the Participating Institutions, the National Science Foundation, the U.S. Department of Energy, the National Aeronautics and Space Administration, the Japanese Monbukagakusho, the Max Planck Society, and the Higher Education Funding Council for England. The SDSS Web Site is http://www.sdss.org/.

The SDSS is managed by the Astrophysical Research Consortium for the Participating Institutions. The Participating Institutions are the American Museum of Natural History, Astrophysical Institute Potsdam, University of Basel, University of Cambridge, Case Western Reserve University, University of Chicago, Drexel University, Fermilab, the Institute for Advanced Study, the Japan Participation Group, Johns Hopkins University, the Joint Institute for Nuclear Astrophysics, the Kavli Institute for Particle Astrophysics and Cosmology, the Korean Scientist Group, the Chinese Academy of Sciences (LAMOST), Los Alamos National Laboratory, the Max-Planck-Institute for Astronomy (MPIA), the Max-Planck-Institute for Astrophysics (MPA), New Mexico State University, Ohio State University, University of Pittsburgh, University of Portsmouth, Princeton University, the United States Naval Observatory, and the University of Washington.
\bibliographystyle{apj}
\bibliography{bibliography}
	
\end{document}